\documentclass[journal=esthag,manuscript=article,letter]{achemso}
\usepackage{fullpage}
\usepackage[left]{lineno}
\usepackage{hyperref}
\usepackage{dcolumn}
\hypersetup{colorlinks=true,urlcolor=blue,linkcolor=blue,citecolor=blue}
\SectionNumbersOff
\usepackage{amsmath}
\usepackage{mathptmx}
\usepackage{graphicx}
\usepackage{nicefrac}
\title{\fontsize{16}{16}\selectfont Seeing the Forest for the Trees: Scaling Analysis of Energy in Great Lakes Water Supplies} 
\author{\fontsize{14}{14}\selectfont Likwan Cheng}
\affiliation{City Colleges of Chicago, 30 E.\ Lake Street, Chicago, Illinois 60601, United States}
\email{lcheng6@ccc.edu}
\begin{document}\vspace{-16pt}
\centerline{\sffamily ORCID: \href{https://orcid.org/0000-0002-7183-1157}{0000-0002-7183-1157}}
\centerline{Version: October 3, 2019}
\vskip 48pt
{This document is the unedited Author's version of a Submitted Work that was subsequently revised and accepted for publication after peer review in {\it Environmental Science \& Technology}, copyright \copyright\ by American Chemical Society. To access the final edited and Published Work, see \url{https://pubs.acs.org/doi/10.1021/acs.est.9b05982}.
}
\newpage
\begin{abstract}
Resource-scale quantification of energy in water supplies is a basis for regional-scale governance of water systems and national- and global-scale assessments of the ``water-energy-emission" nexus. But a physically based approach for this quantification remains lacking. Here, applying recently emerged complex system principles, we develop ``scaling analysis" (SA), a complex network-based methodology for quantifying energies and energy-nexus properties in water supplies at the resource or regional scale. Conceiving water supply systems as networks embedded in regionally self-organizing complex systems, SA explores the complex system laws of skewed or scaled size abundance (decreasing system abundance with increasing system size) and allometric energy scaling (decreasing system energy intensity with increasing system size) as unifying formulation for profiling system size distribution and predicting system energies from system sizes. Overcoming challenges facing traditional approaches from transscale and unbounded system size distributions and scarcity of energy data, SA represents physically based, data-driven predictions. We empirically demonstrate SA and test its predictions for the water supply systems of the Great Lakes, one of the world's most geographically expansive water resources.
\end{abstract}

\newpage
\section{Introduction}
\begin{figure}[]
\includegraphics[width=0.72\textwidth]{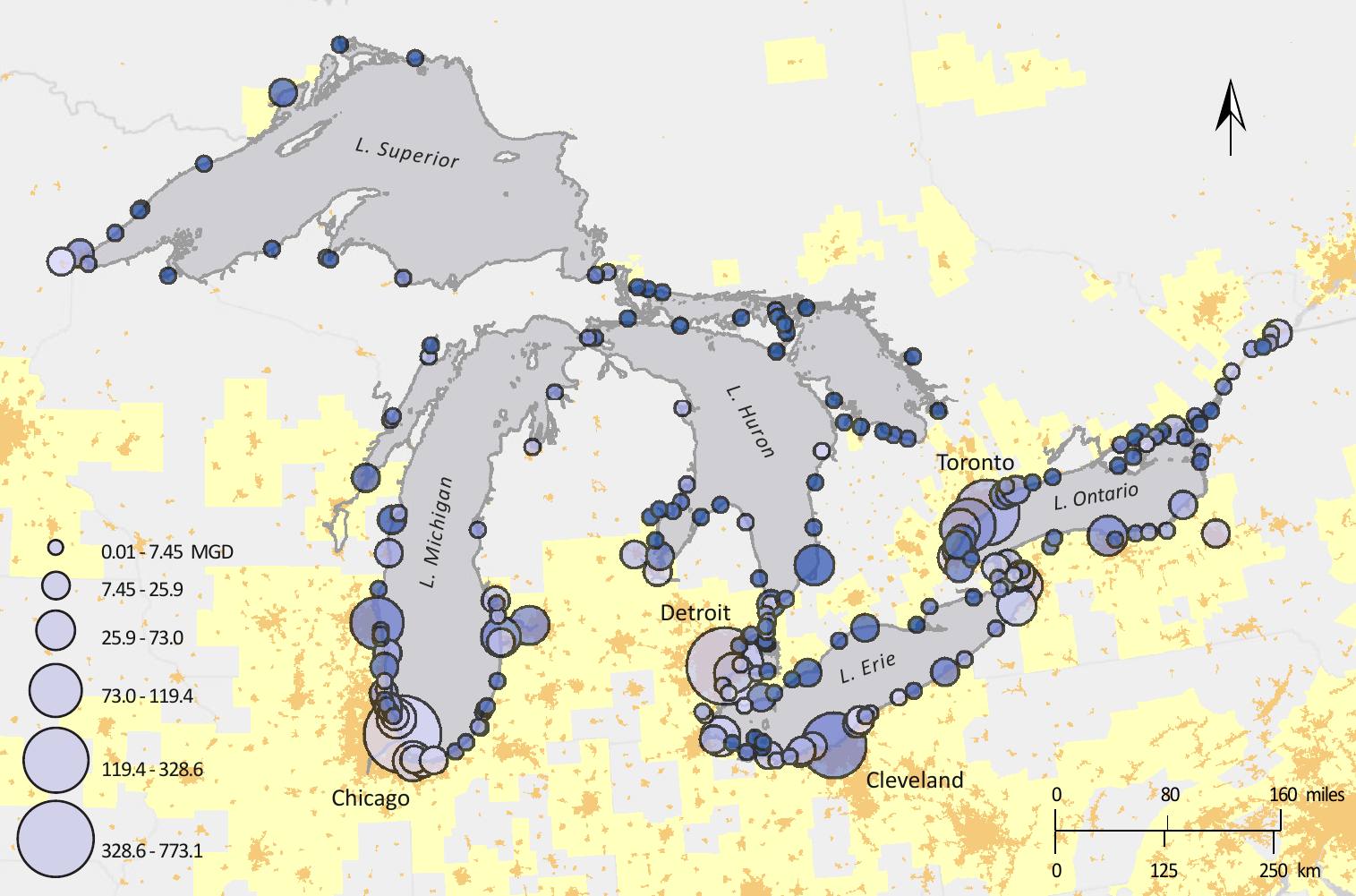}
\caption{\label{gl} Great Lakes public water supplies (circles; $n = 222$) connect one vast homogeneous water resource to a large number ($n \simeq 1\,018$) of highly heterogeneous human communities. Light-brown polygons indicate census-designated metropolitan and micropolitan statistical areas (in the US) or metropolitan areas and agglomerations (in Canada). Dark-brown areas indicate census-designated urban areas (in the US) or population centers (in Canada)\cite{uscen,staca}. Darker blues indicate higher energy intensities.} 
\end{figure}

The Great Lakes are the largest bodies of surface freshwaters on Earth. Among their many vital ecosystem services, the Great Lakes are a resource of drinking water for a tenth of the US-Canadian population. Over 200 public water supply systems draw water from the Great Lakes and provide it to over 1\,000 communities---cities, villages, or rural population centers---along the over 10\,000-mile shorelines (Figure \ref{gl}). The great heterogeneity of these supply systems contrast dramatically the vast homogeneity of the water source. From the small village plants on the North Shore of Lake Superior that serve less than two hundred people to the giant metropolitan systems on the southern rims of Lakes Michigan and Huron that serve many millions---including the world's largest water treatment plant at Chicago, the daily pumpage of the Great Lakes supplies rises through 5 orders of magnitude while their electricity intensities drop by more than half. From the abundances of hydropower on the river banks in New York and Ontario to the richness of coals in the basins of Pennsylvania and Illinois, the energy that powers the Great Lakes supplies carries greenhouse gas emission intensities that vary by as much as a factor of 7. Are there universal orders in these great variations that can be discovered to unify all supply systems of the Great Lakes as a whole? 

\subsection{Challenges in resource-scale energy quantification: No typical system size and few energy data}
Holistic governance of water supplies is best done at the scale of the resource region, because the specificity of the resource context often entails a homogeneity in the supply systems operating within it \cite{her15}. Energy is an important aspect of this governance \cite{doe14}. However, a physically based approach for quantifying supply energy at the resource scale remains lacking. Comprehensive studies of energy for water supplies must comprise two complementary components: an ``inter-region, categorical study" that characterizes energy for dissimilar-type systems in different resource regions, and an ``intra-region, quantitative study" that quantifies energy for similar-type systems within a resource region. So far, researchers have focused on inter-region, categorical studies. Using various approaches, including cross-regional or national surveys \cite{epr09,epr13,two11,sow17,chi18}, region-specific case studies \cite{fil04,rac07,wmo10,sto11,fan15,jeo15,lee17,sto17,xue19}, and inter-region case comparisons \cite{wmo14}, studies have shown that different types of systems have different energy demands---notably in treatment (e.g., freshwater treatment vs.\ desalination) and transport (e.g., local distribution vs.\ trans-basin conveyance). Thus, across resource regions, ``system-type" heterogeneity leads to heterogeneity in energy intensity \cite{gao11}.  

By contrast, intra-region, quantitative studies have seen little advance. Within a resource region, despite system-type homogeneity, supplies vary greatly in size, and ``system-size" heterogeneity also leads to heterogeneity in energy intensity. As a result, resource-scale, intra-region energy quantification faces large challenges. Presently, lacking a formalism to handle these heterogeneities, large-scale energy quantification resorts to the estimation approach of dividing systems into several size classes \cite{epa18} and then using  ``sampling" to seek the ``typical" system energy for each size class \cite{epr13,liu16}. But this class-based, sampling approach is both crude and fundamentally susceptible to large errors. As explained in Figure S1 ({\it Supporting Information}), sampling is only accurate for ``fixed-scale, bounded" distributions. But sizes of a region of water supplies have a ``transscale, unbounded" distribution; such a distribution contains no ``typical" size and its unboundedness means sampling errors can be arbitrarily large\cite{bar}. Instead of sampling, one must use a ``systematic" approach to assemble the sizes of all individual systems in the region and then seek a unifying analytical formulation to relate sizes to energies, system by system.

Recent advances in complex systems have revealed two nonlinear, ``scaled" phenomena. First, complex systems exhibit ``skewed or scaled size abundance" describable by the lognormal \cite{mit11} or the power-law \cite{cla09} distribution function, respectively. Skewed (or scaled) size abundance has been observed in the ensembles of many natural and technoeconomic systems, including  cities \cite{bat08}, corporations \cite{axt01}, river basins \cite{rin14}, terrestrial animals \cite{cla09} and plants \cite{per19}, and aquatic autotrophs \cite{per19}. Second, supply networks embedded in complex systems exhibit ``allometric energy scaling" \cite{wes17}---decreases in energy intensity (energy per volume) with increases in system size. Allometric energy (or metabolic) scaling has been observed in the ensembles of many natural and living systems, including animals \cite{wes97}, plants \cite{vas18}, ecosystems \cite{nid18}, and the runoff \cite{rin14} and vegetation \cite{rod11} systems in river basins.

Most recently, the phenomena of skewed (and scaled) size abundance and allometric energy scaling were simultaneously observed for the first time in the anthropogenic ensemble of public water supplies (of the US state of Wisconsin) \cite{che17}. These discoveries revealed that sizes and energy intensities in water supplies do not assume arbitrary distributions or values, but follow complex system laws. At the ensemble (resource or region) level, the size distribution follows the ``laws of abundance" \cite{bat13}---the larger the system size, the less abundant the system. At the individual (supply system) level, energy intensities follow the ``law of allometry" \cite{bar17}---the larger the system size, the lower the energy intensity. The observed allometric energy scaling was quantitatively explained by a network-based theory \cite{che17}. Further temporal observations revealed that water supply networks evolve by following a ``common evolutionary track" \cite{che18}, so that the energy scaling is ``universal", to be expected for all supply systems with approximately constant scaling parameters. 

\subsection{Scaling analysis: a complex network approach for energy quantification}
These complex system laws provided the unifying analytical formulation needed to relate system sizes to system energies. In this paper, we introduce ``scaling analysis" (SA), a complex network-based methodology for quantifying energy in water supplies at the resource scale. In its essence, SA represents physically based, data-driven predictions. SA is a multi-level approach (Table 1): first, from top down, it bases on the law of abundance to guide the discovery of individual system sizes; second, at the system level, it bases on the law of allometry to predict system energies from system sizes; finally, from bottom up, it sums the system energies to give the ensemble energy, allowing one to ``{\it see} the forest for (because of) the trees." By taking this route, SA circumvents the theoretical challenge of ensemble transscale nonlinearity and unboundedness (no typical size) and the practical challenge of scarcity of individual energy data. Here, using a recently completed census data on Great Lakes water supplies \cite{che19}, we empirically demonstrate SA and statistically test the robustness of its predications in the realistic geospatial context of one of the world's most expansive water resources. 

\newpage
\section{Methods}
\subsection{Definitions}\label{sec_def}
We study the public water supply (PWS) systems of the Great Lakes surface water (GLSW), the water in the Great Lakes or their connecting waterways \cite{glc13}. GLSW as a regional resource subdivides into five subregions, each defined by one Great Lake and any downstream waterways \cite{che19}: (1) Lake Superior and St.\ Mary's River; (2) Lake Michigan; (3) Lake Huron, St.\ Clair's River, St.\ Clair's Lake, and Detroit River; (4) Lake Erie and Niagara Rivers; and (5) Lake Ontario and the portion of St.\ Lawrence River up to the New York-Ontario border limit. The Great Lakes reside within the jurisdictions of the US states of Minnesota, Wisconsin, Illinois, Michigan, Indiana, Ohio, Pennsylvania, New York, and the Canadian province of Ontario.

A PWS system provides water for residential, commercial, institutional, and most industrial consumption in a community. We define ``supply system" (SS) as a PWS system that draws from GLSW, treats the water with (usually) one water treatment plant (WTP), and distributes the treated water to consumers. We define ``consumption community" (CC) as a place where the water is consumed---a city, village, town or a population center within a town, water district, or unincorporated census place. We define the ``size" of an SS or CC according to its annual flow quantity, measured in units of million gallons per day (MGD). Usually, SS sizes are based on withdrawal flows from the source (in US jurisdictions) or raw-water flows through the WTP (in Ontario), whereas CC sizes are based on consumption flows. For GLSW, we identified a total of 222 active SSs, which operate a total of 235 WTPs (2013 figures) and serve a total of 1018 CCs \cite{che19}.

The physical structure of an SS is dominated by the distribution network, the hierarchically ordered web of pipelines spanning from the WTP. The typical SS comprises one WTP and serves its own CC plus perhaps one or a few neighboring CCs. An SS can also be operated jointly by several neighboring CCs. The large metropolitan SS comprises multiple WTPs and serves the host city plus a large number of suburban CCs. Such a multiple-WTP system is considered as a single SS by virtue of the overall morphology of its urban distribution network. In Canada, some SSs are operated by the upper-tier municipalities of regions; each regional SS has its own WTP and serves a distinct group of CCs.

The energy of an SS is defined as its overall annual electricity use, including the phases of water intake, treatment, and transmission and distribution. For the conventional systems at the Great Lakes (no inter-basin conveyance or desalination), this energy is dominated by the transmission and distribution phase, through centralized high-lift pumping from the WTP or distributed pump-station pumping throughout the network. Often, a metropolitan SS provides only transmission pumping to distant suburban CCs, where additional pumping by local supplemental ``secondary systems" distributes the water to consumers. This additional level of distribution hierarchy reflects the self-organization of the large urban network. With secondary systems, the SS energy no longer represents the entire ``source-to-tap" supply energy. For perspective, the rural Alexandria Bay SS (0.17 MGD) has one WTP and serves its own village. The suburban Evanston SS (38.9 MGD) has one WTP and serves its own and a neighboring city, plus four distant suburban CCs supplemented with secondary systems. The metropolitan Chicago SS (773 MGD) has two WTPs and 12 pump stations and serves the city of Chicago plus some 125 suburban CCs, many of which are supplemented with secondary systems \cite{che19}.

\subsection{Data}
Data, data sources, and acquisition approaches are presented in a separate document \cite{che19}. In summary, the data were collected from original and official sources, including public water utilities, municipal governments, and jurisdiction regulatory authorities. Since laws of the Great Lakes jurisdictions vary, multiple approaches were used in data collection. The data comprise four types of datasets: SS water flows, SS energies, CC infrastructural (transmission and distribution network) lengths and volumes, and CC sizes. 

For the identification of Great Lakes PWS systems and the quantities of their annual flows, jurisdiction-level centralized databases were used for the US jurisdictions and system-level reports were used for Ontario. For energy (electricity use), jurisdiction-level centralized databases were available only for Wisconsin and Ontario. For systems in other jurisdictions, the electricity data were obtained from one of the following sources: (1) official water utility annual reports, (2) direct requests made to superintendents of water utilities, (3) public record requests made to municipal governments, and (4) jurisdiction audit reports of local governments. Systems operated by private entities ($n = 4$) are not included for the energy portion of this study for lack of data. Electricity data for very small, island-based plants located on northern Lake Huron that use electricity in place of natural gas for space heating ($n=8$) were subjected to an adjustment by a reduction by 20\%. For infrastructure, data were  collected only for municipalities that host an SS and usually from governmental reports; these partial data proved to be sufficient for the supporting purpose of demonstrating infrastructural scaling. For CC sizes, data were obtained from various documented sources or estimated in proportion of populations. The data for SS size, SS energy, and CC infrastructural length and volume are 100\%, 96\%, 53\%, and 23\% complete, respectively.

\subsection{Size abundance}
We consider the size abundance of the SS and CC ensembles. Like city size distributions \cite{luc17}, the SS and CC distributions are best described by the lognormal distribution in the body and the power law in the upper (large-size) tail. We fit the whole distribution with the lognormal model and the upper tail with the power-law model.

The lognormal distribution is described by the probability density function (PDF) 
\begin{equation}
p = \frac{1}{\sqrt{2\pi}\sigma M} \exp[-\frac{(\ln M-\mu)^2}{2\sigma^2}],
\end{equation}
where $M$ is the size variable, and the lognormal parameters $\mu$ and $\sigma$ are the mean and the standard deviation (SD), respectively, of the associated log-transformed normal distribution \cite{mit11}. The density profile is characterized by the median $e^\mu$ and the skewness $(e^{\sigma^2}+2)\sqrt{e^{\sigma^2}-1}$. 
Following standard practice \cite{cla09,mit11}, we fit the data as empirical cumulative distribution function (CDF) in log-log space rather than as PDF in linear space to avoid inaccuracies associated with histogram binning. We then display the data and fit as complementary cumulative distribution function (CCDF) in log-log space for critical visualization. 

The power-law distribution is described by the PDF 
\begin{equation}
p\propto M^{-\alpha},
\end{equation}
where $-\alpha $ is the exponent. In standard practice, the power law is also analyzed in log-log space where $-\alpha$ is the slope of a linear function. Since in the present case power law appears only at the upper tail, we use the discrete generalized beta distribution (DGBD) function to fit the dataset in rank-size distribution \cite{xli15}. The DGBD function is \cite{mar09}
\begin{equation}
M(r) \propto \frac{1}{r^\alpha}(n+1-r)^\beta, 
\end{equation}
where $r$ is the rank of size $M$, $n$ the length of ranks, and $\alpha $ (power-law exponent) and $\beta $ are fitting parameters. The DGBD function reduces to a pure power law when $\beta =0$ and further reduces to Zipf's rank-size law $M(r) \propto r^{-1}$ when $\alpha =1$. Zipf's rank-size law describes the size distributions of large cities \cite{bat13}. 

\subsection{Allometric scaling}
Scaling describes the the nonlinear relation between a property, such as energy $E$, and size $M$ in an ensemble. Scaling is commonly modeled as a power law

\begin{equation}
E = AM^b,
\end{equation}
where $A$ is the prefactor and $b$ is the exponent; for allometric scaling, $b< 1$. In practice, this nonlinear relation is analyzed as a linear relation in log-log space after a logarithmic transformation, 

\begin{equation}
\log E = a + b\log M, 
\end{equation}
where $a= \log A$ is the intercept and $b$ is the slope of the linear model.

The mechanistic basis of allometric energy scaling in water supplies lies at the infrastructural transport network \cite{che17}. The infrastructural network exhibits ``spatial" and ``structural" scaling \cite{bet13}. A water supply network comprises transmission and distribution mains. Given the dominance of the transmission and distribution phase in energy use, which may account up to 85\% of the overall system energy \cite{pla12,epr09}, the energy of the entire system can be approximated as that of the transport network.

Spatial scaling arises from a dimensional relation in community growth---a three-dimension spanning consumer mass resting on a two-dimension spanning land area. The network length $L$ (area-filling) scales sublinearly with community size $M$ (mass-filling) to the two-third power, or $L \sim M^\frac{2}{3}$ \cite{bet13,che17}. Network structural scaling arises from the fact that the network evolves self-organized hierarchical orders through the merging of mains. Main merging causes network volume $V$ grows faster than network length $L$, so that $V$ scales superlinearly with $L$ to the five-quarter power, or $V \sim L^\frac{5}{4}$ \cite{bet13,che18}.

Infrastructural scaling effectuates a shortening (from the sublinear scaling of $L$ with $M$) and widening (from the superlinear scaling of $V$ with $L$) of the distribution network on per-unit size basis, implying decreased energy intensities with increased network (or system) sizes. Based on the Hazen-Williams energy dissipation force \cite{kar00} on a hierarchical infrastructural network model \cite{bet13}, the energy scaling exponent was derived to be allometric, with $b=0.82$ \cite{che17}. The intercept $a=\log A$ depends only on spacial and structural parameters, and can be determined empirically by considering systems of sizes $\log M = 0$. From systems within $\log M =0\pm 0.02$ MGD ($n = 6$), we obtained an empirical intercept $a=5.88$ (SD, 0.16). The parameterized energy scaling law is

\begin{equation}\label{eq_e}
E = 10^{5.88} M^{0.82} \quad (\rm {kWh}),
\end{equation}
where $M$ is in units of MGD. If the parameters $(a,b)$ assume invariant values across  ensembles from region to region, the energy scaling law is said to be ``universal", as is believed to be in organisms \cite{wes17}. For water supplies, the argument for universality rests on infrastructural optimization (which determines $b$) and consumer homogeneity (which selects $a$) \cite{che18}. 

\newpage
\section{Data-driven discovery of scaling laws}
\subsection{Size abundance skewness and scaling}
The SS dataset ($n=222$) totals an annual flow of 3\,600 MGD \cite{che19}, in agreement with the GLSW withdrawal data for public supplies (3\,593 MGD) by the official Great Lakes Commission (GLC) (adjusted for St.\ Lawrence River withdrawal) \cite{glc13}. The breakdowns of the SS annual flow by jurisdiction and by lake are also in agreement with GLC data (average percentage difference are 4.1\% and 1.6\%, respectively) \cite{che19}. These comparisons confirm the completeness and accuracy of the present dataset. Great Lakes public supply withdrawals \cite{glc13} account for a significant 8.8\% of the combined public supply withdrawals by the US (39\,000 MGD \cite{die18}) and Canada (2\,630 MGD \cite{ec11}). The CC dataset ($n=1018$) totals an annual flow of 3\,240 MGD. The CC figure is 12\% below the SS figure, a difference in line with water loss in distribution.

\begin{figure}
\includegraphics[scale=0.9]{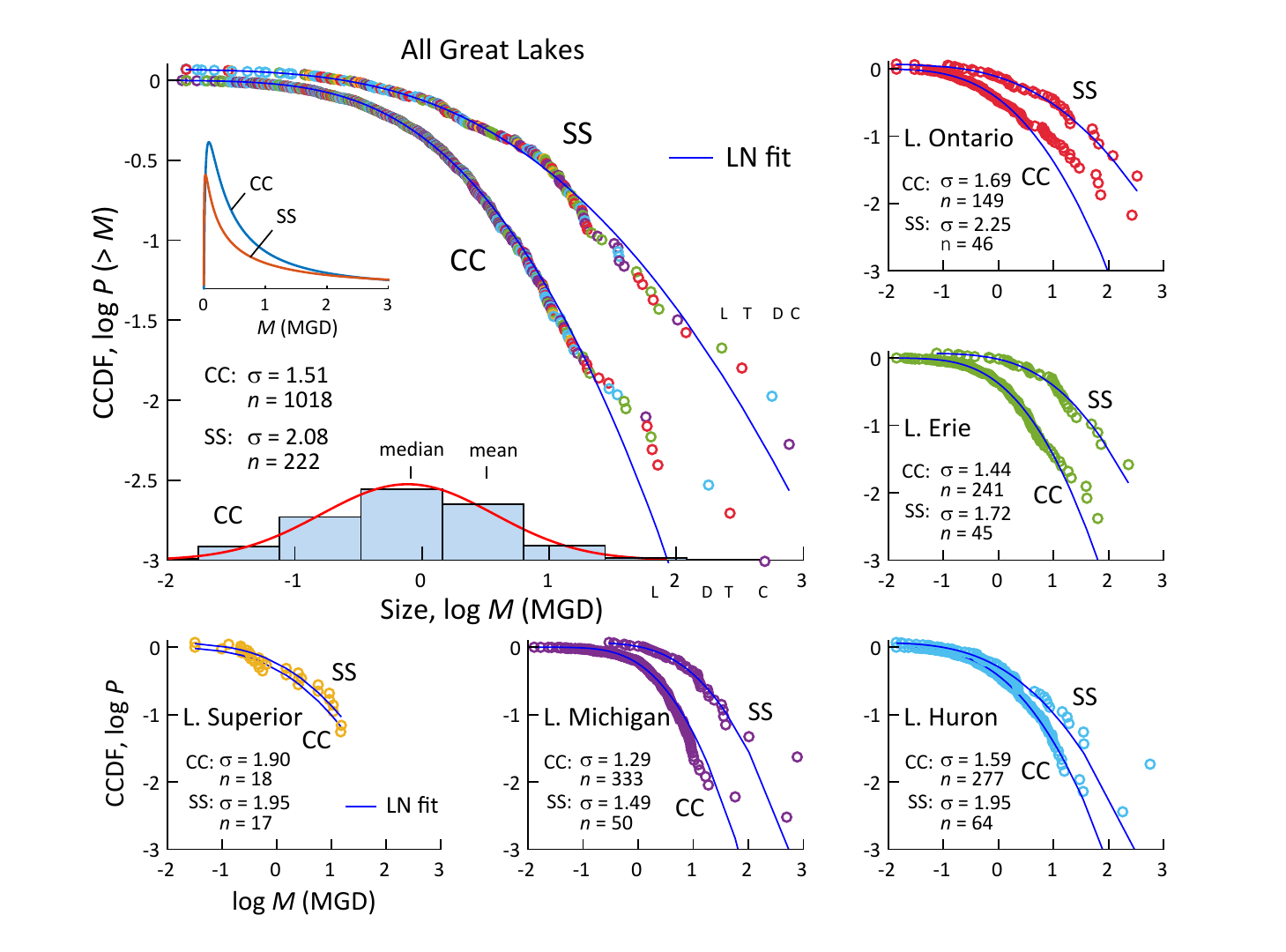}
\caption{\label{lgn} Lognormal distributions of supply system (SS) and consumption community (CC) sizes. SS and CC sizes (circles, color-coded by lake) are plotted as empirical complementary cumulative distribution function (CCDF) in log-log (10-based) space for all Great Lakes (large panel) and for individual lakes (small panels), with lognormal distribution best-fits (blue curves). In the all-Great Lakes panel, inset plot shows fitted SS and CC size distributions in linear-linear space; histogram shows CC size distribution as log-transformed empirical probability density function (PDF) with normal distribution fit (red curve); letters indicate the four largest SSs and their CCs: Chicago (C), Detroit (D), Toronto (T), Cleveland (L). For display clarity, the SS curves are slightly shifted on the vertical axis. The logarithm is 10-based.}
\end{figure}

We first examine the lognormal fit for the body of the distribution. Figure \ref{lgn} shows the CC and SS data and their respective lognormal distribution fits as CCDF for both the all-Great Lakes and the individual-lake data. We focus on the all-Great Lakes data. The smooth CC data points allowed fitting refinements by excluding a small portion of lower-tail to optimize the fit, with the best-fit estimates of $\mu = -0.21$ and $\sigma=1.51$. The SS data yields a satisfactory lognormal fit with the best-fit estimates of $\mu = 0.75$ and $\sigma=2.08$.

We next examine the power-law fit for the upper tail of the distribution. Figure \ref{pwl} shows the CC and SS data and their respective DGBD function fits as rank-size distributions. We again focus on the all-Great Lakes data. The straight lines of data points at the upper tail suggest the presence of power law. The observed power-law exponents for the CC and the SS datasets are $\alpha = 0.87$ and $\alpha = 1.14$, respectively.

The ensemble transition from CC to SS represents a community aggregation accompanied by an increase in both lognormal $\mu$ (indicator of ensemble median) and $\sigma$ (indicator of ensemble skewness) as well as an increase in power-law $\alpha$ (indicator of size disparity). Consistently, $\mu$ and $\sigma$ of the Great Lakes SS ensemble are notably larger than those of the Wisconsin groundwater SS ensemble ($\mu = -2.03$, $\sigma = 1.64$) (Table 2). Both of the above observations can be explained by a ``large gets larger" phenomenon in network aggregation, and this favoring for the large is consistent with property allometry.

For the lognormal body, the Great Lakes CC ensemble has a skewness ($\sigma = 1.51$) comparable to that of Swiss municipalities ($\sigma=1.36$) \cite{dec07}, whereas the Great Lakes SS ensemble has a skewness ($\sigma = 2.08$) greater than that of US cities and places ($\sigma \simeq 1.75$) \cite{luc17}. These comparisons concur with expectations, as Great Lakes CCs are often rural and include smaller units (e.g., districts) but the largest SSs are larger than cities. For the power-law tail, the Great Lakes SS ensemble has an exponent ($\alpha = 1.14$) comparable to that of US urban clusters ($\alpha = 1.22$) as measured as nighttime lightness \cite{xli15}. This comparison also concurs with expectation as both water supplies and light clusters reflect spatially aggregated intensities of human activities.

\subsection{Allometric energy scaling}
We first describe the results of spatial and structural scaling as bases of energy scaling. Figure S2 shows that the network length $L$ scales with community size $M$ with an exponent of $0.71(2)$, in agreement with the $\frac{2}{3}$-power expectation. Figure S3 shows that the network volume $V$ scales with network length $L$ with an exponent of $1.19(2)$, in agreement with the $\frac{5}{4}$-power expectation. The slight deviations in both observations are explainable as urban densification effects and were previously observed \cite{che17}.

\begin{figure}
\includegraphics[scale=0.9]{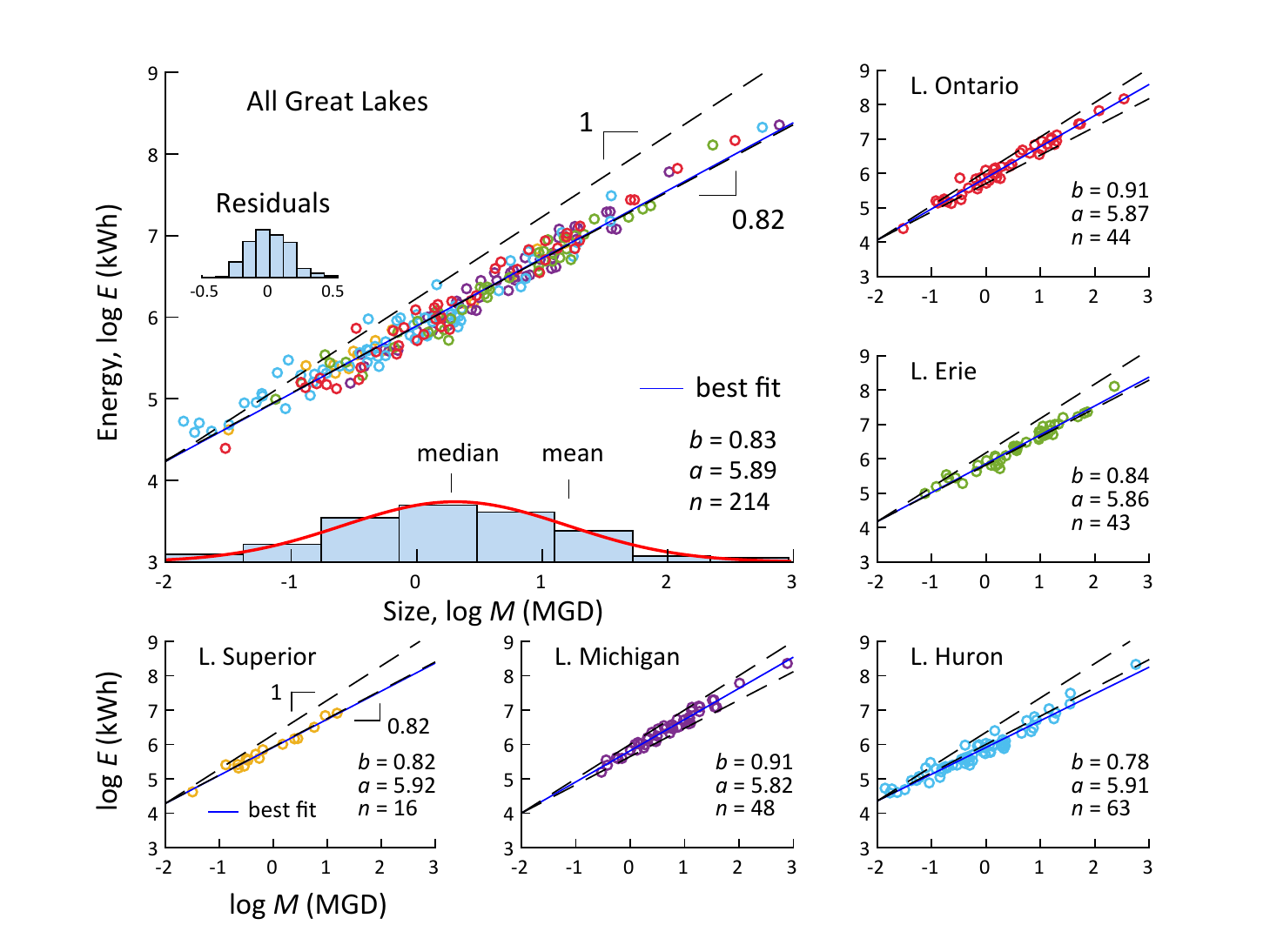}
\caption{\label{kwh} Allometric energy scaling in SSs. SS annual energies $E$ (circles) are plotted against sizes $M$ in log-log space for all Great Lakes and for individual lakes, with linear best-fits (blue lines). Black dashed-lines indicate the theoretical slope ($b= 0.82$) and isometric slope ($b= 1$), respectively. In the all-Great Lakes panel, inset shows the normally distributed fit residuals; histogram shows the SS size distribution as log-transformed empirical PDF with normal distribution fit (red curve).}
\end{figure}

We now describe the result of energy scaling. Figure \ref{kwh} shows that, for the all-Great Lakes dataset, system energy $E$ scales with system size $M$ with an exponent $b = 0.83(2)$ and intercept $a = 5.89(2)$, in agreement with scaling law-expectation, $(a,b)= (5.88,0.82)$, and with the Wisconsin groundwater result, $(a,b)= (5.82,0.85)$ \cite{che17}. These parametric agreements support the notion of universal scaling. The individual-lake observations are generally consistent with the all-Great Lakes observation. 

The residuals of the best fit is normally distributed (Kolmogorov-Smirov test for standard normal, $p=0.90$) with an SD of $0.15$ (Figure \ref{kwh}). The variability comes primarily from ``inter-system" variations \cite{che18}, attributable to evolved system adaptations to local contexts, such as network topology (urban gridiron vs.\ rural branching), landscape topography (leveled vs.\ hilly), consumer geography (centralized vs.\ dispersed), and treatment technology (conventional vs.\ membrane). Besides, the variability also contains ``intra-system" variations, attributable to random year-to-year variations within each system. This latter variation is quantified to have an SD of $0.05$ based on the time series of the energy vs.\ flow data for available supplies (Figure S4).

We may express the scaling in units of energy intensity $e$:

\begin{equation}\label{eq_int}
e = \frac{E}{365M} = 2143 \times M^{-0.17} \quad (\rm {kWh/MG}),
\end{equation} 
where $M$ is in units of MGD. The negative slope of $-0.17$ explicitly expresses decreases in per-unit energy demands with increases in system size (Figure \ref{prd}a). Several ensemble statistics are notable. The expected energy intensity of the median-size system (1.89 MGD) is 2\,113 kWh$/$MG. The expected energy intensity of the mean-size system (16.27 MGD) is 1\,495 kWh$/$MG. The large differences among these statistics underscore the effects of skewness and allometry, concurring with that there is no single representative system for the ensemble.

\subsection{Deviation analysis of scaling parameters}\label{sec_geospa}
The simplicity of the complex system laws allows the ensemble to be described with only a few parameters---$(a,b)$ in allometry law, $\alpha$ in power law, and $(\mu,\sigma)$ in lognormal equation. However, complex system laws differ from the classical laws of physics in that they represent the statistical propensities of self-organization and adaptation \cite{ott04}, which are inherently contextually dependent. The ``analysis" in SA emphasizes proper interpretation of parametric deviations. Here, we give geospatial and contextual explanations for the deviations in the power-law and the allometric exponents. 

\begin{figure}
\includegraphics[scale=0.9]{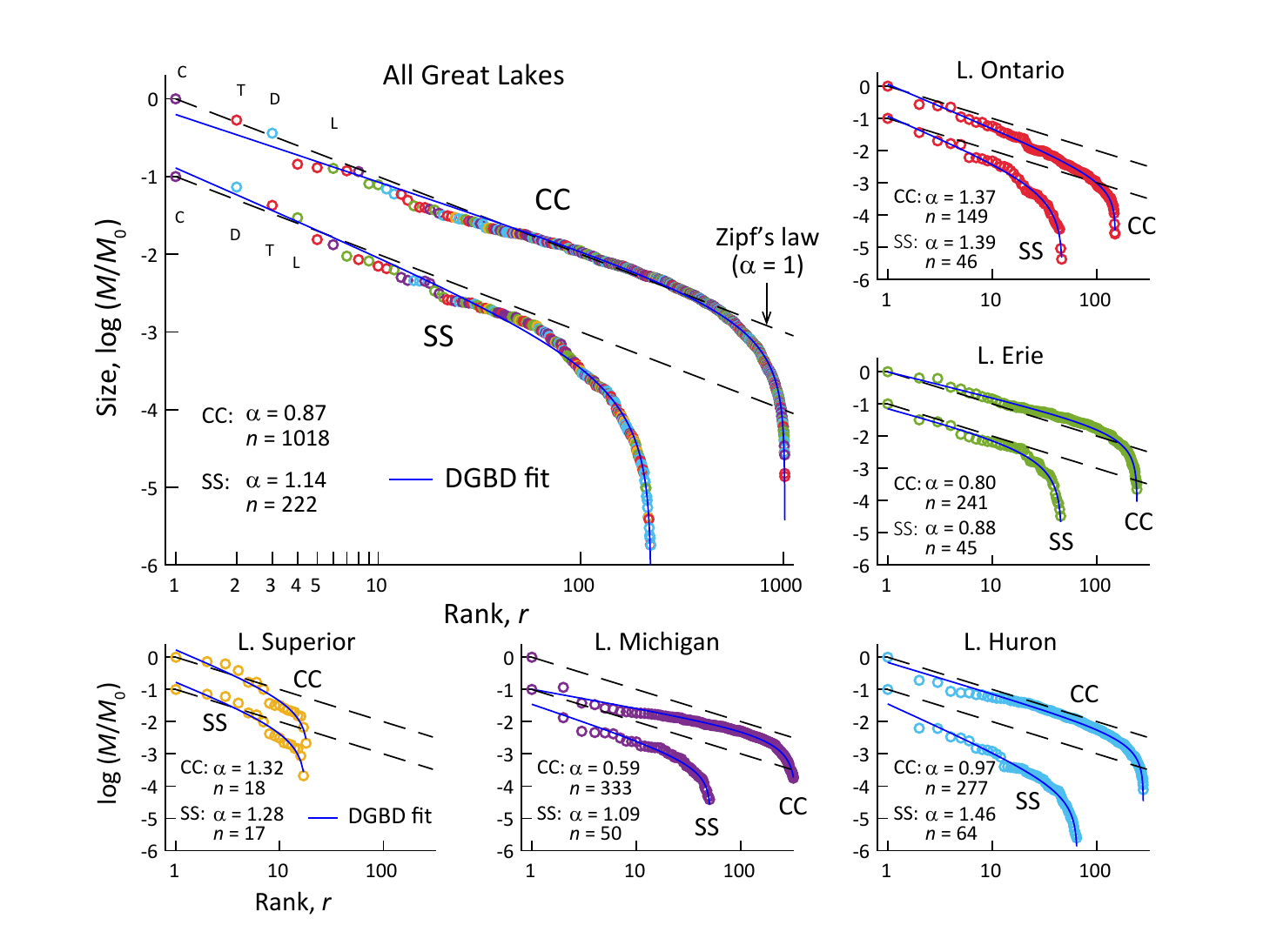}
\caption{\label{pwl} Power-law distributions of the upper tails of SS and CC sizes. SS and CC sizes (circles) are plotted as empirical rank-size distribution in log-log space for all Great Lakes and for individual lakes, with discrete generalized beta distribution (DGBD) best-fits (blue curves). Black dashed-lines indicate Zipf's law. For display clarity, the SS curves are shifted on the vertical axis.}
\end{figure}

Scaled size abundance in the form of Zipf's law exists in the spatial domain of ``ensemble completeness" \cite{cri12}. This domain has an ``inward bound", below which incomplete inclusion of members may lead to scaling deviations. The all-Great Lakes SS dataset shows a power-law at the upper tail close to Zipf's law (Figure \ref{pwl}), indicating the all-Great Lakes region constitutes a complete ensemble. Now, if we shrink the spatial domain from the all-Great Lakes region to an individual-lake region, the accompanying deletions of ensemble members will result in reductions in both size range and rank range. In particular, deleting non-leading data points (reducing rank range but not size range) results in discontinuities (jumps) in the upper tail and increased $\alpha $ (Lake Michigan and Lake Huron); deleting the leading data point(s) (reducing size range more than reducing rank range) results in decreased $\alpha $ but without jumps in the upper tail (Lake Erie and Lake Ontario). This explains the observed deviations in the individual-lake results (Figure \ref{pwl}).

Allometric energy scaling exists within the spatial domain of ``system-type homogeneity" \cite{che18}. This domain has an ``outward bound", beyond which system-type heterogeneity may lead to scaling deviations. Consider energy (electricity), the intra-region system homogeneity over the entire Great Lakes region explains the observed consistent $(a,b)$ parameters for both the all-Great Lakes dataset and the individual-lake datasets (Figure \ref{kwh}). By contrast, consider greenhouse gas (GHG) emissions associated with the life cycle of the electricity, $G = \epsilon E = \epsilon AM^b$, where $\epsilon$ is the effective emission intensity factor \cite{che17}. Since $\epsilon$ varies considerably by jurisdiction---from $ 0.86$ kgCO$_2$e$/$kWh in Indiana \cite{eia12} to $ 0.11$ kgCO$_2$e$/$kWh in Ontario \cite{oeb13}---the spatial domain of system homogeneity is the jurisdiction. As a result, $G$ scales with $M$ within individual or similar jurisdictions with jurisdiction-dependent intercepts $a=\log A + \log \epsilon $, but $G$ does not scale with $M$ for the all-Great Lakes dataset (Figure S5). 

Besides, local contextual variations may also perturb scaling exponents. For example, Lake Huron, which hosts mostly small rural systems (median, 0.72 MGD) in colder climate, has a lower exponent ($b= 0.78$), whereas Lakes Michigan and Ontario, which host many larger suburban systems (median, 5.74 MGD and 1.57 MGD, respectively) in more temperate climate, have a higher exponent ($b= 0.91$ for both). The local adaptation explanation of scaling deviations \cite{che18} finds supports in other scaled systems. In size abundance, deviations in $\alpha $ were observed in the global scaling of national economies \cite{cri12} and the scaling of phytoplankton populations across the global ocean \cite{per19}. In energy allometry, deviations in $b$ were observed in the scaling of metabolic rates in organisms across body sizes \cite{kol10}.

\newpage
\section{Scaling law-driven energy predictions}\label{sec_EngPrd}
The purpose of SA is to make predictions: for individual systems whose energy data are unavailable. Here, we evaluate the robustness of the predictions for individual and ensemble energies by comparing empirical and predicted data.

\subsection{Individual energies}
We test the predictability of individual system energies $E(M)$. We randomly select a system from the ensemble and generate an ``empirical" energy data point for this system according to the empirical value but allow for intra-system variation (SD, 0.05). We next generate a ``predicted" energy data point for the same system using the empirical scaling law but allow for inter-supply variation (SD, 0.15). We then evaluate the absolute fractional difference between the empirical and the predicted values. We repeat this selection and evaluation procedure for a large number of times and calculate the cumulative average absolute difference, which converges at 0.17 on the logarithmic scale, indicating a reliable predictability (Figure S6). 

\subsection{Ensemble energy}
\begin{figure}
\includegraphics[scale=0.9]{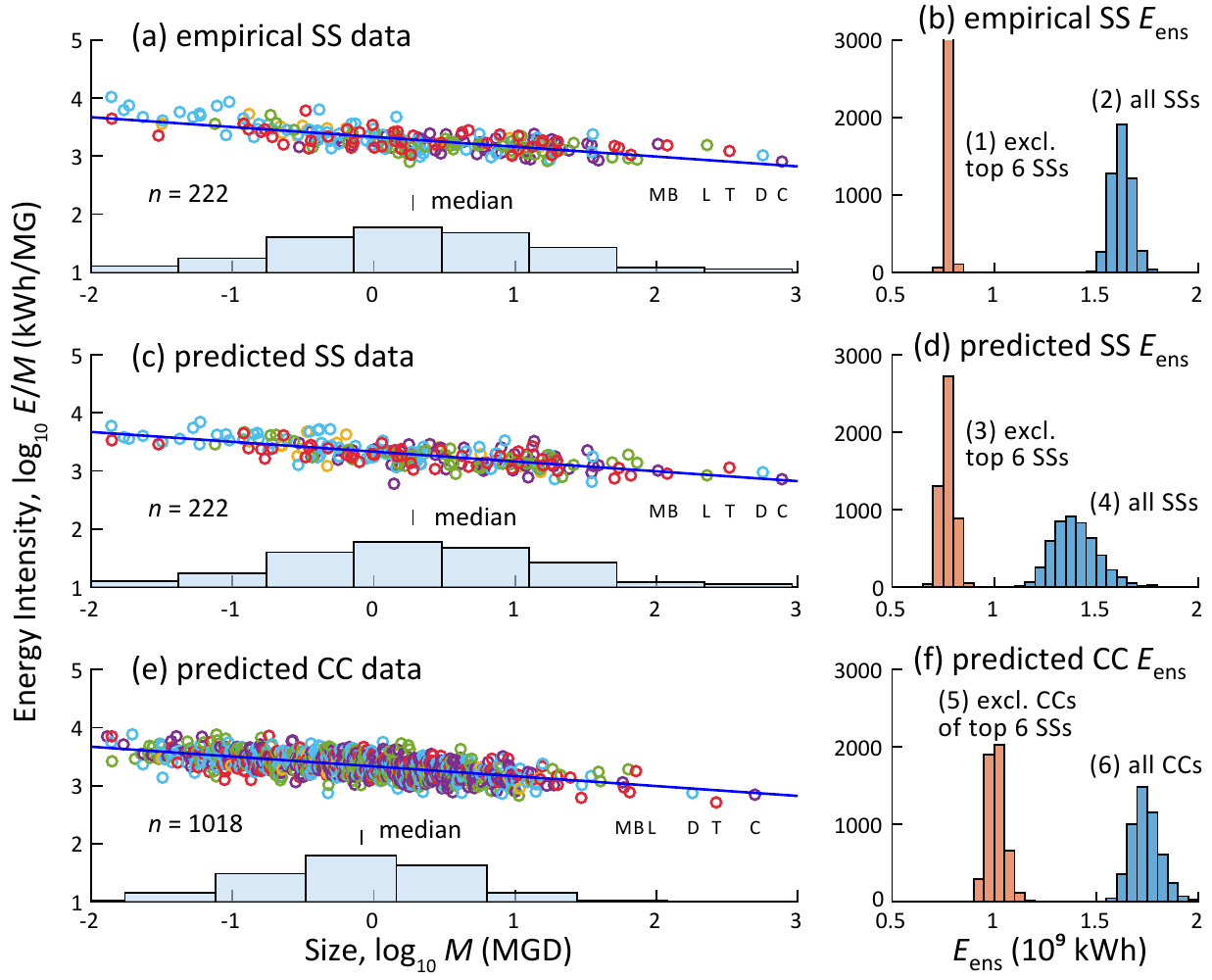}
\caption{\label{prd} Empirical and predicted individual and ensemble energies. (a) empirical SS individual energy intensities (circles; missing data points filled according to scaling law) with scaling law (blue line); (b) histograms of empirical SS ensemble energies with (blue) and without (orange) the largest six SSs; (c) predicted SS individual energy intensities; (d) histograms of predicted SS ensemble energies with (blue) and without (orange) the largest six SSs; (e) predicted CC individual energy intensities; (f) histograms of predicted CC ensemble energies with (blue) and without (orange) the CCs supplied by the largest six SSs. Letters in (a,c,e) indicate the largest six SSs: Chicago (C), Detroit (D), Toronto (T), Cleveland (L), Mississauga/Brampton (B), and Milwaukee (M).}
\end{figure}

We test the predictability of the ensemble energy $E_{\rm ens}$. We generate ``empirical" and ``predicted" energy data as before but now for the entire ensemble ($n =222$) (Figure \ref{prd}a,b). We repeat this process for a large number of dataset pairs. We first test the predictability of the SS ensemble energy. To avoid the deviation effects in the largest systems, we do the test with the largest six systems excluded---Chicago, Detroit, Toronto, Cleveland, Mississauga/Brampton, and Milwaukee. We sum the datasets for the empirical and the predicted $E_{\rm ens}$, respectively (Figure \ref{prd}b,d; peaks 1,\,3). The empirical $E_{\rm ens}$ is $0.78 (0.75,0.80) \times 10^9$ kWh annually (95\% confidence interval in parenthesis) and the predicted $E_{\rm ens}$ is $0.77 (0.71,0.83) \times 10^9$ kWh. This agreement (two-sample $t$-test, $p=0.51$) confirms the predictability of the scaling law. Second, we evaluate the effect of the largest systems by repeating the procedure with the largest six systems included. Now the empirical $E_{\rm ens}$ is $1.63 (1.53,1.72)\times 10^9$ kWh and the predicted $E_{\rm ens}$ is $1.40 (1.20,1.62)\times 10^9$ kWh (Figure \ref{prd}b,d; peaks 2,\,4). The discrepancy reflects the effect of deviations in the largest six systems from scaling law predictions (Figure \ref{prd}a). Third, we attempt to predict the ensemble source-to-tap energy by including the energies of the secondary systems. To do this, we use the same evaluation procedure but now base it on the CC dataset (Figure \ref{prd}e). For the subset of CCs not served by the largest six systems $(n = 659)$, the predicted $E_{\rm ens}$ is $1.01 (0.93,1.09) \times 10^9$ kWh (Figure \ref{prd}f, peak 5); for all CC of the  ensemble $(n = 1018)$, the predicted $E_{\rm ens}$ is $1.74 (1.61,1.88) \times 10^9$ kWh (Figure \ref{prd}f, peak 6). From the all-CC result, the secondary system energy can be roughly estimated to be $ 0.34 \times 10^9$ kWh (difference between peaks 6 and 4).

\newpage
\section{Policy applications}
We consider two broad lines of applications inspired by scaling analysis. First, with its multi-level perspective, SA helps holistic governance by connecting urban-level (local) policy actions with multiurban-level (regional) impact assessments, allowing using maximizing regional impact as guide for decisions on local actions. For example, consider urban stormwater capture to augment potable supply \cite{lut19}. For energy reduction, are these capture actions better implemented in small towns (to target higher energy intensities) or in large cities (to target higher potable flows)? This question may be answered by an analysis of scaling exponents. In river scaling, drainage-basin areas scale sublinearly ($b\simeq \nicefrac{2}{3}$) with runoff volumes \cite{rod11}, just like urban area scales sublinearly ($b\simeq 0.71$) with potable volumes \cite{che17}. In urban scaling, urban impervious surfaces scale slightly superlinearly ($b\simeq 1.1$) with urban areas \cite{qma18}. These relations suggest that urban runoffs are likely to scale superlinearly with urban areas. Here, because the spatial exponent ($b\simeq \nicefrac{2}{3}$) is more allometric than the energy exponent ($b \simeq 0.82$), it favors placing capture systems in large cities over small towns for greatest effects. Similar considerations may be applied to assess other issues related to the distribution network, such as water loss \cite{col02}.

Second, with the allometric scaling law, SA may be incorporated as a data-driven predicted tool into policy decisions. These decisions may include making forecasts \cite{ath17}; appropriating resources \cite{riv18}; planning for infrastructure renewals \cite{aww12}; setting regulatory benchmarks \cite{bee12}; setting water valuation \cite{rog10} and pricing \cite{olm09}; adapting water supply to urbanization \cite{lar16} and quantifying its roles in urban economies \cite{kru96}; and articulating the local vs.\ centralized debate \cite{she12} and the built (gray) infrastructure vs.\ green infrastructure debate \cite{mul15}. Specifically for the Great Lakes, SA may help translate the Great Lakes Compact \cite{glc08} from guiding principles of environmental protection to quantifiable policy metrics.


\begin{suppinfo}
Figure S1: Comparison of bounded and unbounded distributions.\\
Figure S2: Scaling of network length with community size. \\
Figure S3: Scaling of network volume with network length.\\
Figure S4: Time series of scaling of system energy with system size for selected Great Lakes water supplies.\\
Figure S5: Predicted scaling of electricity-associated GHG emissions with system size.\\
Figure S6: Traces of cumulative average absolute difference between predicted and observed energies of supply systems.\\
Figure S7: Traces of cumulative average values of the ensemble energies under various conditions.\\
\end{suppinfo}

\newpage
\begin{table}[]
\centering
\caption{Schematic steps of scaling analysis.}\label{table1}
\footnotesize
\setlength\tabcolsep{6pt}
\begin{tabular}{cll}
\hline\hline\rule{-2.5pt}{3ex}
steps   &  procedures   &  quantities \\[1ex]
\hline\rule{-2.5pt}{3ex}
1 & Abundance law-guided systematic assembly of individual sizes  & $p(x)$ \\
2 & Allometry law-based predictions of individual energies from individual sizes  & $E(x) = Ax^b$ \\
3 & Obtaining ensemble energy by summing individual energies & $E_{\rm ens} = \int p(x)E(x)dx$ \\[1ex]
\hline
\end{tabular}
\end{table}

\newpage
\begin{table*}[]
\centering
\caption{Comparison between Great Lakes surface water and Wisconsin groundwater supply ensembles in size distribution, infrastructural scaling, and energy scaling.}\label{table2}
\footnotesize
\setlength\tabcolsep{4pt}
\begin{tabular}{lccccccccc}
\hline\hline\rule{-2.5pt}{3ex}  
supply ensembles                              &        &       \multicolumn{3}{c}{size distribution} &  & \multicolumn{4}{c}{scaling}     \\
\cline{3-5} \cline{7-10}\rule{0pt}{3ex}                                                                      
                                             &        &   \multicolumn{2}{c}{lognormal} & power law &  model  & \multicolumn{2}{c}{intercept $ a=\log A$}  &  \multicolumn{2}{c}{slope $b$}     \\
                                           & $n$ & $\mu $   & $\sigma $ & $\alpha$        &                & expected  &   observed    &  expected &  observed                 \\[1ex]
\hline\rule{-2.5pt}{3ex}
Great Lakes surface water & 222 & 0.75 & 2.08 &         1.14               & $L=AM^b$   & --- & 5.44(4)  & $\nicefrac{2}{3}$ & 0.72(4)   \\ 
                   &       &         &         &                                 & $V=AL^b$   & --- & 4.55(6) & $\nicefrac{5}{4}$ & 1.20(5)   \\ 
                                            &       &         &         &                                & $E=AM^b$  &  5.88 & 5.89(2)  & $0.82$ & 0.83(2)    \\[2ex] 
Wisconsin groundwater & 518 & $-2.03$ & 1.64 & 1.01  & $L=AM^b$   & ---  & 5.43(3)  & $\nicefrac{2}{3}$ & 0.71(2)    \\ 
                                    &       &         &         &           & $V=AL^b$   &  ---  & 4.54(2) & $\nicefrac{5}{4}$ & 1.19(2)   \\  
                                                                 &       &         &         &           & $E=AM^b$   & 5.88  & 5.82(3)  & 0.82 & 0.85(3)    \\[1ex] 
\hline                                   
\end{tabular}
\end{table*}

\newpage

\newpage

\newpage

\newpage

\newpage


\begin{thebibliography}{99}
\bibitem{che19} Cheng, L. ``Sizes, electricity uses, and community consumptions of Great Lakes water supplies". (Note: This data document will be posted with the publication of the paper.)
\end{thebibliography}
\end{document}


\vspace{-14pt}
\centerline{\sffamily ORCID: \href{https://orcid.org/0000-0002-7183-1157}{0000-0002-7183-1157}}
\centerline{Version: October 3, 2019}

\begin{figure}[]
\centering
\centerline{\includegraphics[width=0.9\textwidth]{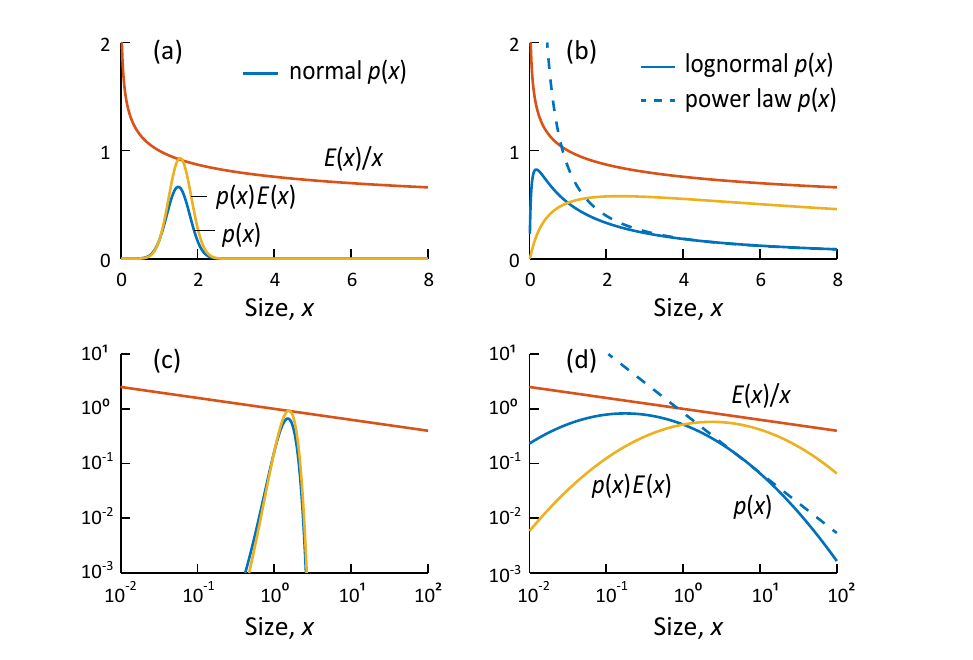}}
\caption{\label{norm} Comparison between (a,c) the normal distribution and (b,d) the lognormal and power-law distributions of a random size variable $x$, shown in (a,b) linear-linear space and (c,d) log-log space, respectively. $p(x)$ (blue) is the probability density function of the normal (SD, 0.3), lognormal ($\sigma = 1.5$), and power-law ($\alpha = 1.1$) distributions, respectively; $E(x)/x$ (red) is the energy intensity function, where $E(x)$ is the allometric energy function ($b = 0.8$); and $p(x)E(x)$ (yellow) is the ensemble energy at size $x$. For the normal distribution, $p(x)$ is exponentially ``bounded". A randomly selected $x$ falls within a narrow range of the ensemble mean $\bar x$ because the standard deviation is of the same scale as the mean \cite{bar}. The bounded nature of the normal distribution allows $\bar x$ to be used as an approximation for the PDF and $p(x)E(x) \sim p(\bar x)E(\bar x)$ is insensitive to the variation of $E(x)$. Thus, the evaluation of the ensemble energy can be simplified to random sampling because of the convergence of the sampled $x$ to $\bar x$. For a power-law or power-law-tailed lognormal distribution, $p(x)$ is ``unbounded". A randomly sampled $x$ can be arbitrarily large because the standard deviation of the power-law distribution is divergent \cite{bar}. The unbounded nature of the power-law implies that there is no typical $x$ sampling can capture. The evaluation of $p(x)E(x)$ must be done by analytically integrating $E(x)$ over $p(x)$, or by numerically summing $E(x)$ for all $x$, because $E(x)$ is a transscale nonlinear (power-law) function in itself.} 
\end{figure}

\begin{figure}[]
\centering
\includegraphics[width=0.9\textwidth]{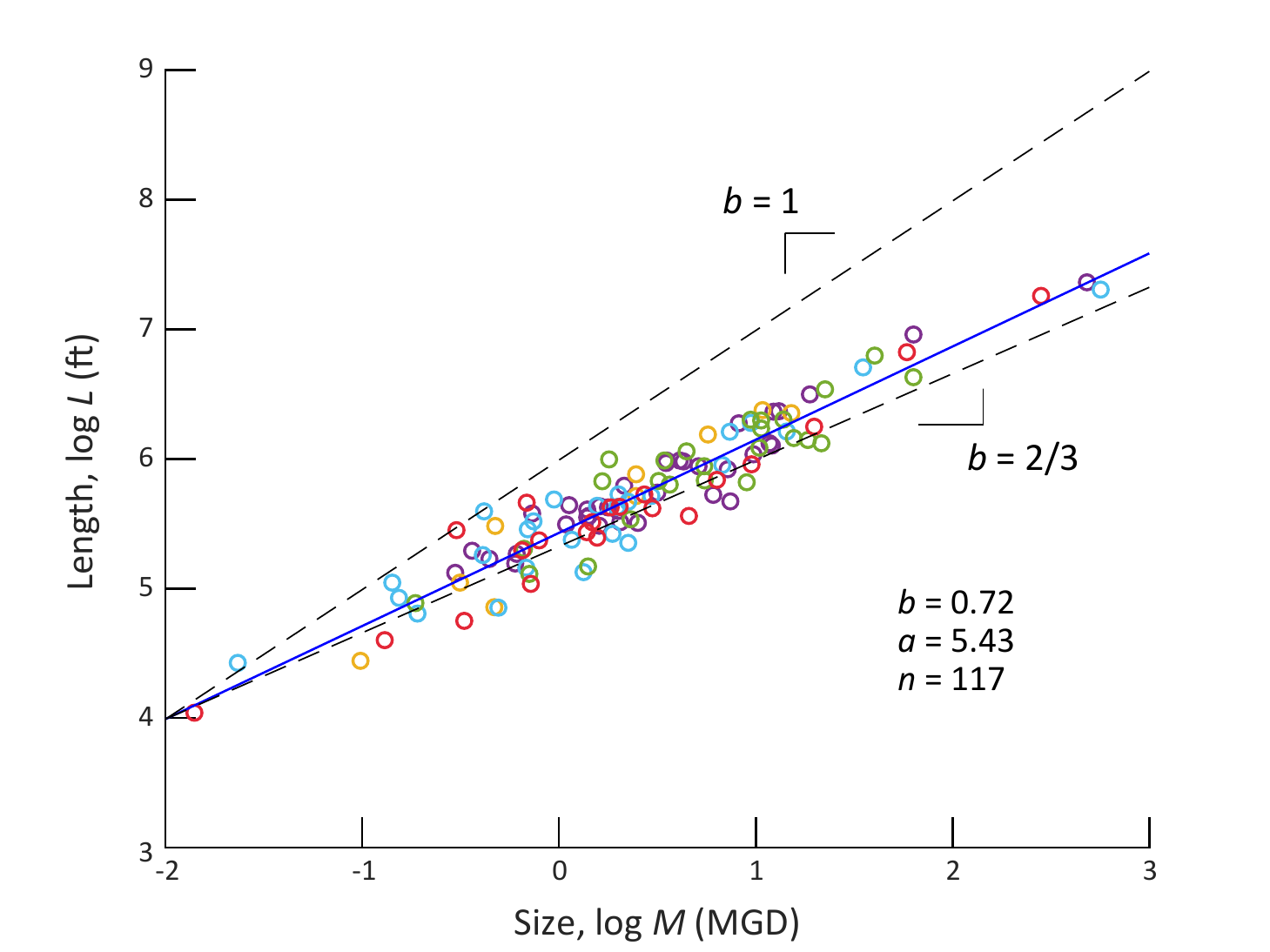}
\caption{\label{lin} Scaling of network length $L$ with size $M$ for selected Great Lake communities that host a supply system\cite{che19}, with linear best-fit (blue line). Black dashed-lines indicate theoretical scaling ($b=\frac{2}{3}$) and isometric scaling ($b =1$), respectively. Data points are color-coded by lake as indicated in Figure 4 in the main text. The logarithm is 10-based.}
\end{figure}

\begin{figure}[]
\centering
\includegraphics[width=0.9\textwidth]{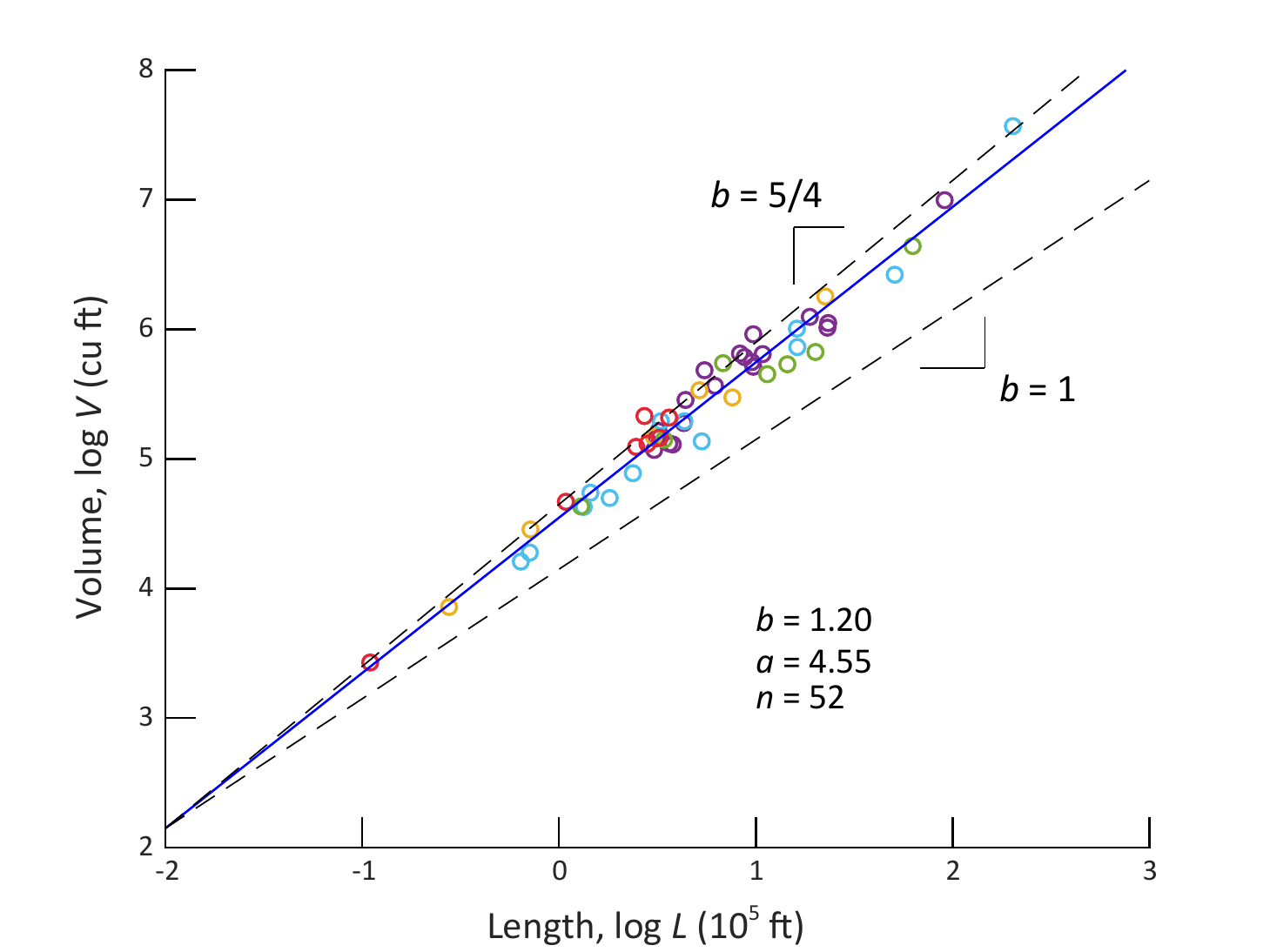}
\caption{\label{vol} Scaling of network volume $V$ with network length $L$ for selected Great Lake communities that host a supply system\cite{che19}, with linear best-fit (blue line). Black dashed-lines indicate theoretical scaling ($b=\frac{5}{4}$) and isometric scaling ($b =1$), respectively.}
\end{figure}

\begin{figure}[]
\centering
\includegraphics[width=0.9\textwidth]{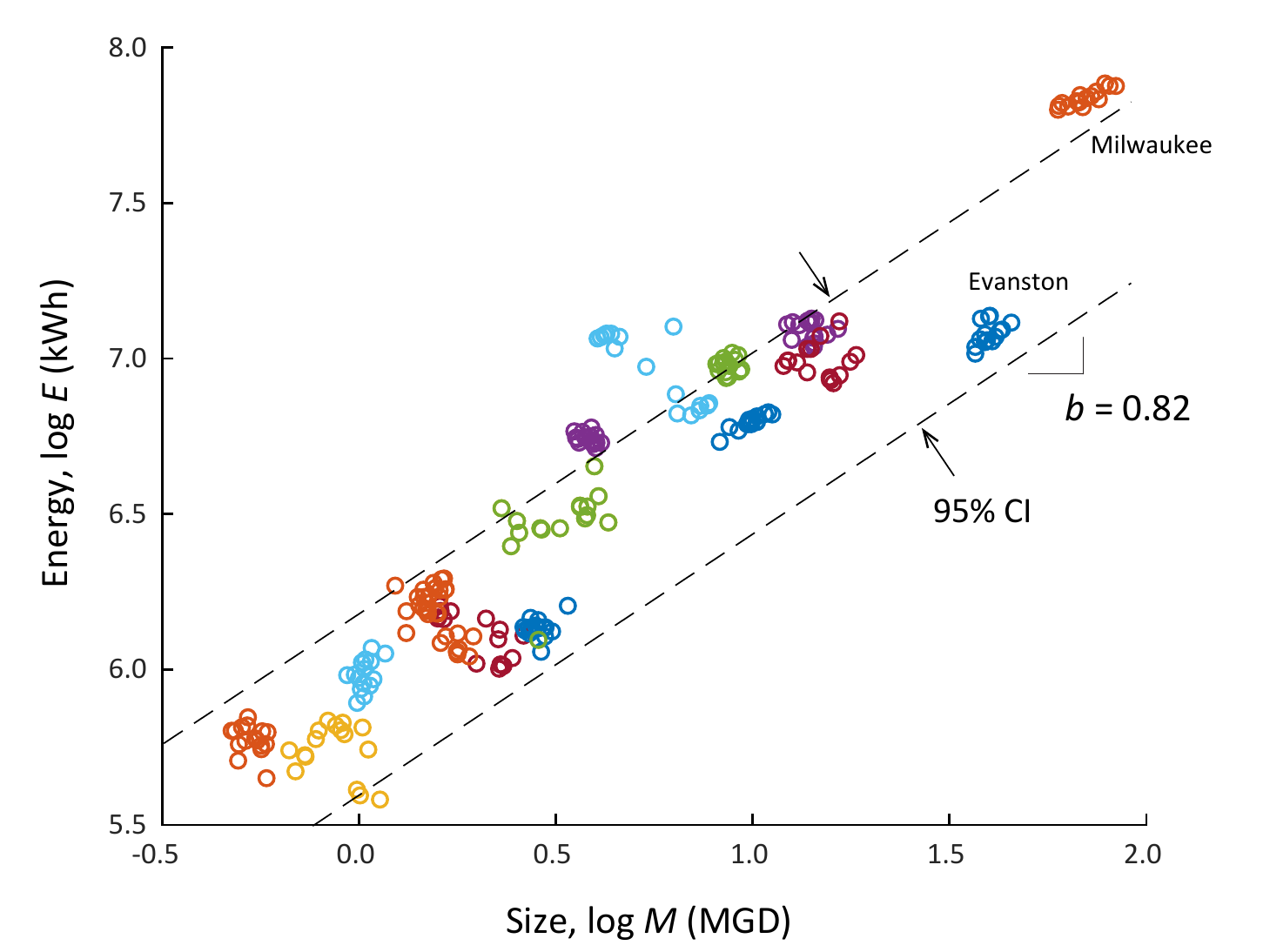}
\caption{\label{ind} Scaling of system energy (electricity) with system size as time series for 15 available supply systems (each shown in a distinct color) and for a period of 20 years (1997-2016). Dashed-lines indicate the 95\% confidence interval of the residual distribution from the fitted line for energy scaling, representing ``inter-system variations" (SD, 0.15). An ``intra-system variation" (SD, 0.05) is calculated from the vertical variations of each individual system then averaged over all systems. The data are drawn from the systems at Ashland\cite{Ashland}, Superior\cite{Superior}, Cudahy\cite{Cudahy}, Green Bay \cite{Green Bay}, Kenosha\cite{Kenosha}, Manitowoc\cite{Manitowoc}, Marinette\cite{Marinette}, Milwaukee\cite{Milwaukee}, Oak Creek\cite{Oak Creek}, Port Washington\cite{Port Washington}, Racine\cite{Racine}, Sheboygan\cite{Sheboygan},South Milwaukee\cite{South Milwaukee}, Two Rivers\cite{Two Rivers}, and Evanston\cite{Evanston}.}
\end{figure}

\begin{figure}[]
\includegraphics[width=0.8\textwidth]{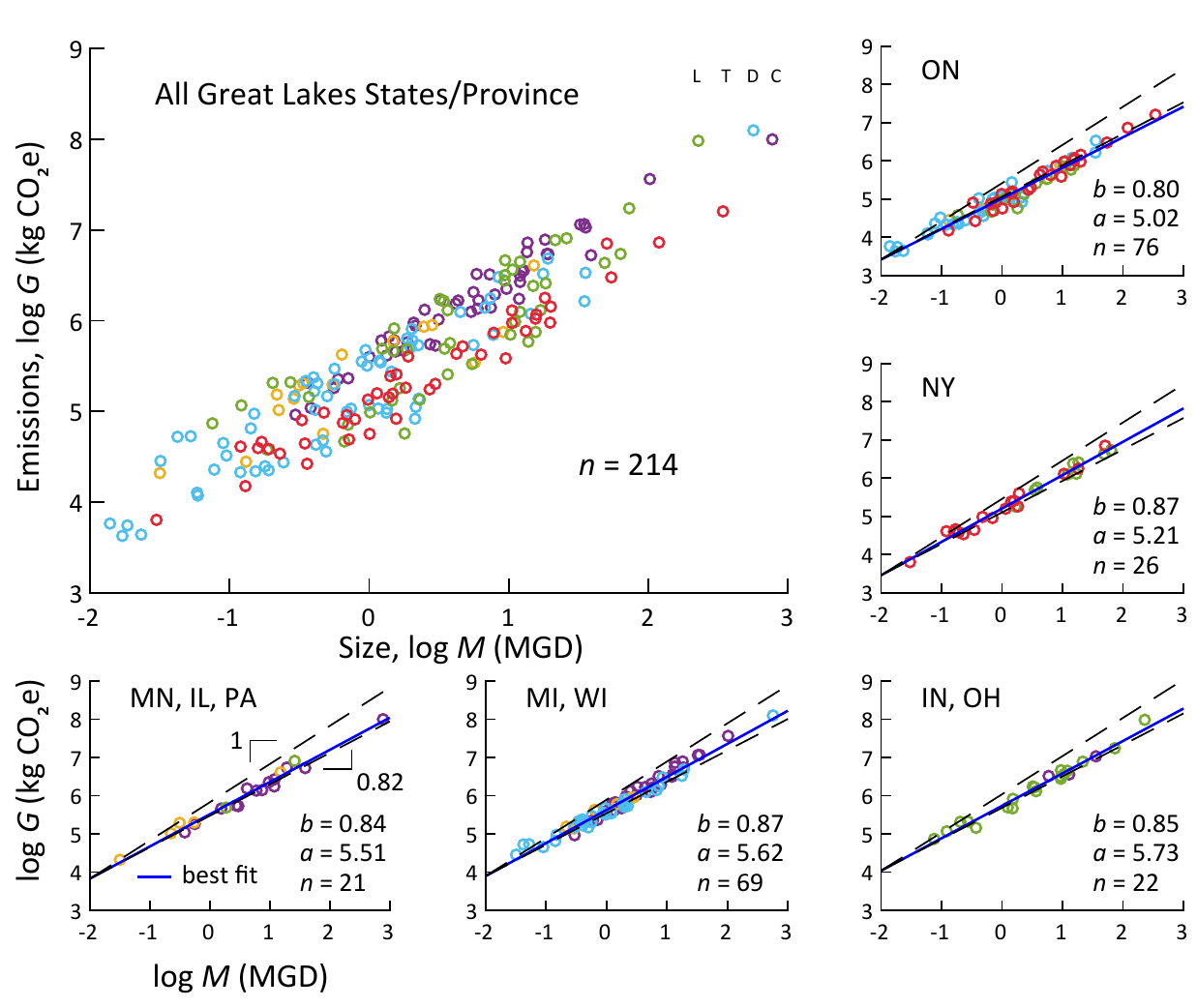}
\caption{\label{ghg} Allometric GHG emission scaling in SSs. Predicted annual GHG emissions $G$ (circles; color-coded by lake) associated with electricity use are plotted against system size $M$ in log-log space for the all-Great Lakes dataset and for individual or similar jurisdictions datasets.} 
\end{figure}

\begin{figure}[]
\centering
\includegraphics[width=0.9\textwidth]{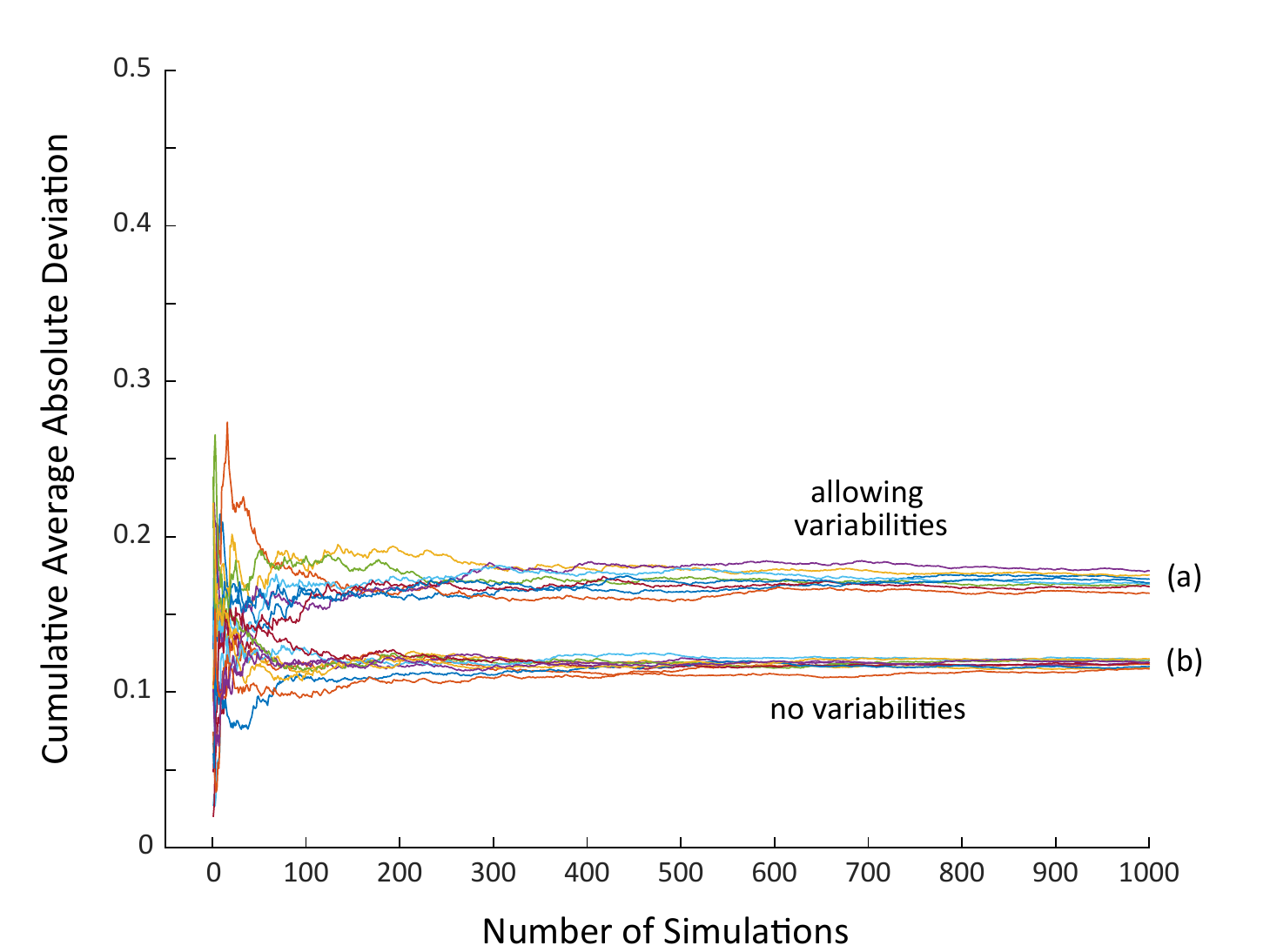}
\caption{\label{} Cumulative average absolute difference in log space between empirical and predicted individual system energies for randomly selected individual systems (a) with and (b) without variabilities. Multiple traces are shown.}
\end{figure}

\begin{figure}[]
\centering
\includegraphics[width=0.9\textwidth]{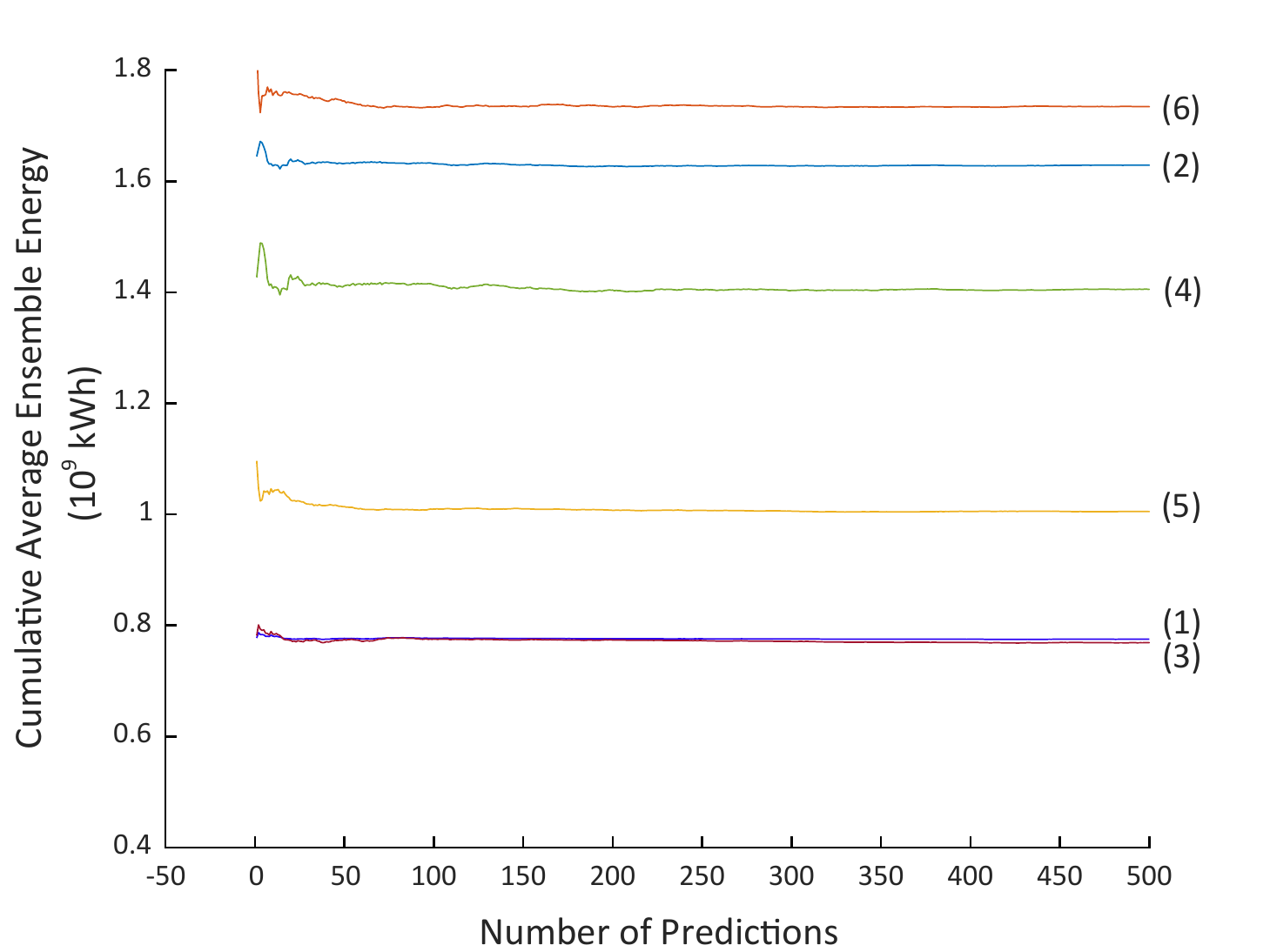}
\caption{\label{mc} Cumulative averages of ensemble energies (trace numbers correspond to peak numbers in Figure 5b,d,f) (1) empirical SS ensemble energy excluding the largest six systems; (2) empirical SS ensemble energy including all systems; (3) predicted SS ensemble energy excluding the largest six systems; (4) predicted SS ensemble energy including all systems; (5) predicted CC ensemble energy excluding CCs served by the largest six systems; and (6) predicted CC ensemble energy including all systems.}
\end{figure}